\documentclass{PoS}

\usepackage{amsmath, amsfonts}
\usepackage{amssymb}
\usepackage{graphicx} 

\title{Electric dipole transitions in pNRQCD}

\ShortTitle{Electric dipole transitions in pNRQCD}

\author{\speaker{Piotr Pietrulewicz}\\
        Faculty of Physics, University of Vienna, Boltzmanngasse 5, A-1090 Wien\\
        E-mail: \email{piotr.pietrulewicz@univie.ac.at}}

\abstract{
We present a theroretical treatment of electric dipole (E1) transitions of heavy quarkonia based on effective field theories. Within the framework of potential nonrelativistic QCD (pNRQCD) we derive the complete set of relativistic corrections at relative order $v^2$ to the decay rate in a systematic, model-independent way. Former results from potential model calculations will be scrutinized and a phenomenological analysis with lattice input in relation to experimental data will be presented.
}

\FullConference{Xth Quark Confinement and the Hadron Spectrum\\
                 8-12 October 2012\\
                 TUM Campus Garching, Munich, Germany}

\begin{document}

\section{Introduction}

Radiative transitions play an important role for our understanding of QCD, in particular of heavy quarkonia. They provide information about the wave functions describing the physical system and probe both the perturbative and non-perturbative regime. Especially E1 transitions give significant contributions to the total decay rate and yield clean signals, which are observed in the experimental facilities. In the last few years CLEO, BES and the B factories have improved their observations of radiative transitions, a review about recent developments can be found in \cite{QWG}.

On the theory side, electric dipole transitions were treated in several potential models, a summary can be found in \cite{Eichten_Godfrey}. We will refer to \cite{Grotch} for comparison with our results. A model-independent treatment to check and improve the calculations has been missing so far. However, in the last decade there has been significant progress using effective field theories (EFTs) to describe heavy quarkonium (see \cite{quarkonium_review} and references therein). Since heavy quarkonium is assumed to be a non-relativistic system we may take advantage of the hierarchy of scales $m \gg mv \gg mv^2$, where $v \ll 1$ is the heavy quark velocity, $m$ is the heavy quark mass ("hard scale"), $p \sim mv$ is the relative momentum of the bound state ("soft scale") and $E \sim mv^2$ is the binding energy ("ultrasoft scale"). The ultimate EFT living at the ultrasoft scale is potential non-relativistic QCD (pNRQCD). In 2005, for the first time radiative decays, concretely 
M1 transitions, were calculated in this theory \cite{M1_Brambilla}. Using the framework of that paper as a guideline we close the remaining gap and compute the decay rates of E1 processes between S and P states (like $n^3 P_J \rightarrow {n'}^3 S_1 \, \gamma$). The following is based on \cite{E1_Brambilla}.

\section{Framework}

By integrating out the hard scale $m \gg \Lambda_{QCD}$ from the fundamental theory (QCD) in perturbation theory ($\alpha_s (m) \ll 1$) one obtains non-relativistic QCD (NRQCD) \cite{NRQCD,Bodwin}. For the calculation of E1 transitions at relative order $v^2$ only the two-fermion Lagrangian $\mathcal{L}_{2-f}$ matters and the relevant part reads 
\begin{equation}
\mathcal{L}_{2-f} = \psi^{\dagger} \left( iD_{0} + \frac{{\bf D}^2}{2m} + \frac{{\bf D}^4}{8m^3} \right) \psi + e e_Q  \psi^{\dagger} \left( \frac{c_F^{em}}{2m} {\bf \boldsymbol \sigma \cdot B}^{em} +i \frac{c_s^{em}}{8m^2} {\bf \boldsymbol \sigma \cdot } [{\bf D} \times,{\bf  E}^{em}] \right) \psi + c.c. \, .
\end{equation}
with $iD_0 = i\partial _0 - g T^a A_0^a - e e_Q A_0^{em}$, $i{\bf D} = i\boldsymbol \nabla + g T^a {\bf A}^a + e e_Q {\bf A}^{em}$ and $\psi$ denoting a Pauli spinor for the heavy quark. The matching coefficients are found to be $c_F^{em} = 1 + C_F \alpha_s(\mu_H)/ 2 \pi + \mathcal{O}(\alpha_s^2) \label{cF}$ and $c_s^{em} = 2 c_F^{em} -1 \label{cs}$ with $\mu_H \sim m$. 

For processes at the ultrasoft scale, NRQCD is not yet the appropriate theory, since there are still several scales entangled ($p, E, \Lambda_{QCD}$) and thus no homogeneous power counting can be established. Integrating out the soft scale $mv$ we obtain a theory for ultrasoft modes, i.e. pNRQCD \cite{pNRQCD:Pineda,pNRQCD:Brambilla}. The crucial step to disentangle the energy and momentum scale is the multipole expansion in the relative distance $r$. To be definite we will use the power counting in the weak-coupling regime ($p \gg E \gtrsim \Lambda_{QCD}$), which reads
\begin{equation}
r \sim 1/mv, \, \boldsymbol \nabla _r \equiv \partial /\partial {\bf r} \sim mv, \, \boldsymbol \nabla \equiv \partial /\partial {\bf R} \sim mv^2, \,{\bf E},  {\bf B} \sim (mv^2)^2, \, {\bf E}^{em} , {\bf B}^{em} \sim k_{\gamma}^2 \label{weak}\, .
\end{equation}
$k_{\gamma}$ is the energy of the emitted photon, which scales like $mv^2$ for transitions between states with different principal quantum numbers.

The pNRQCD-Lagrangian contributing at NLO in the decay rate, i.e. at order $ k^{3}_{\gamma} v^0 / m^2$, reads 

\begin{align}\ \label{L_pNRQCD}
\mathcal{L}_{\textrm{pNRQCD}}  = & \int d^3 r \mathrm{ Tr} \left\{ S^{\dagger} \left( i {\partial}_0 + \frac{{\boldsymbol \nabla}^2}{4m} +\frac{{\boldsymbol \nabla}_r^2}{m}+ \frac{{\boldsymbol \nabla}_r^4}{4m^2} - V_S \right) S +  O^{\dagger} \left( i D_0 + \frac{{\bf D}^2}{4m} +  \frac{{\boldsymbol \nabla}_r^2}{m} - V_O \right) O \right. \nonumber\\
& \hspace{13mm} + \left. g V_A( O^{\dagger} {\bf r}\cdot {\bf E} S + S^{\dagger} {\bf r} \cdot {\bf E} O) \right\} \nonumber \\
&  + \mathcal{L}_{\gamma \textrm{pNRQCD}} + \mathcal{L}_{\textrm{light}}  \, ,
\end{align}
where the covariant derivatives are given by $iD_0 O= i\partial _0 O - g [T^a A_0^a, O]$ and $i{\bf D} O = i\boldsymbol \nabla O + g [T^a {\bf A}^a, O]$ and the trace goes over the color and spin indices. The singlet potential $V_S$ has been calculated perturbatively and non-perturbatively to order $1/m^2$ (\cite{qspectrum, pNRQCD_1m, pNRQCD_1m2}, for more original references see \cite{quarkonium_review}), we display the structure of the relevant potentials for computations at NLO in the decay rate,
\begin{align}
V_S (r) = & \, V^{(0)}(r) + \frac{V^{(1)}_r(r)}{m} + \frac{V^{(2)}_{SI}(r)}{m^2}+ \frac{V^{(2)}_{SD}(r)}{m^2} \, , \label{V1} \\
V^{(2)}_{SI}(r) = & \, V^{(2)}_r(r) + \frac{1}{2} \{ V^{(2)}_{p^2}(r), {\bf p}^2 \} + \frac{V^{(2)}_{L^2}(r)}{r^2} {\bf L}^2 \, , \label{V2} \\
V^{(2)}_{SD}(r) = & V^{(2)}_{LS}(r){\bf L}\cdot {\bf S} +V^{(2)}_{S^2}(r){\bf S}^2 + V^{(2)}_{S_{12}}(r) \left[3 (\hat{\bf r}\cdot \boldsymbol\sigma _1)(\hat{\bf r}\cdot \boldsymbol\sigma _2) - \boldsymbol\sigma _1 \cdot \boldsymbol\sigma _2 \right] \label{V3} \, .
\end{align}
The relevant part of $\mathcal{L}_{\gamma \textrm{pNRQCD}}$ for E1 transitions is 
\begin{align}\label{LagE1}
\mathcal{L}_{\gamma \textrm{pNRQCD}}^{E1} =  e e_Q \int d^3 r \, \mathrm{Tr} & \left\{ V^{r\cdot E} S^{\dagger} {\bf r}\cdot {\bf E}^{em} S + V_O ^{r\cdot E} O^{\dagger} {\bf r}\cdot {\bf E}^{em} O + \frac{1}{24} V^{(r\nabla)^2  r \cdot E}  S^{\dagger} {\bf r}\cdot [({\bf r \boldsymbol \nabla})^2  {\bf E}^{em}] S \right. \nonumber\\
& +i \frac{1}{4m}V^{\nabla \cdot (r \times B)}  S^{\dagger} \{ \boldsymbol \nabla \cdot , {\bf r} \times {\bf B}^{em} \} S \nonumber \\
&+ i \frac{1}{12m} V^{\nabla _r \cdot (r \times (r\nabla)B)}  S^{\dagger} \{ \boldsymbol \nabla _r \cdot , {\bf r} \times  [({\bf r \boldsymbol \nabla}) {\bf B}^{em}] \} S \nonumber \\
& + \frac{1}{4m} V^{(r \nabla) \sigma \cdot B} [ S^{\dagger}, \boldsymbol \sigma ] \cdot [({\bf r \boldsymbol \nabla}) {\bf B}^{em}] S \nonumber \\
& \left. -i \frac{1}{4m^2}V^{\sigma \cdot(E \times \nabla _r)} [ S^{\dagger}, \boldsymbol \sigma] \cdot ({\bf E}^{em} \times \boldsymbol \nabla _{r}) S \right\} \, . 
\end{align}
In fact more terms are allowed according to the symmetries of pNRQCD. However, we can show that their matching coefficients vanish. The matching is done by equating Green's functions in NRQCD and pNRQCD at the energy scale $mv$ order by order in the inverse mass.

\begin{figure}
  \begin{center}
  \includegraphics[width=0.9 \textwidth]{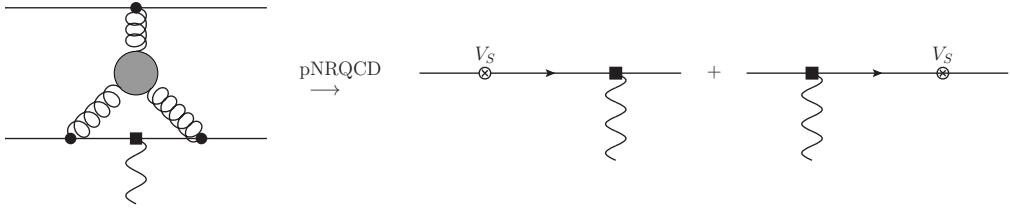}
  \caption{Example for a reducible diagram, if the electromagnetic operator commutes with the gluonic ones. It does not contribute to the matching coefficient of a single operator.}
  \end{center}
  \label{reducible}
\end{figure}

The crucial argument for several operators is that diagrams in NRQCD which can be cast into a reducible structure also give reducible diagrams in pNRQCD. Therefore they have to be subtracted to obtain irreducible operators in pNRQCD and do not play a role in the matching procedure. An example is the diagram in Fig. 1, where the gluonic contribution can be factorized out yielding just a potential.
Using this argument we can fix all of the Wilson coefficients in (\ref{LagE1}) at leading order in $\alpha_{\rm{em}}$, so that the exact results reproduce the ones from tree level calculations, namely
\begin{equation}
V^{r\cdot E} = V_O ^{r\cdot E} = V^{(r\nabla)^2 r \cdot E}= V^{\nabla \cdot (r \times B)}  = V^{\nabla _r \cdot (r \times (r\nabla)B)} = 1, \, V^{(r \nabla) \sigma \cdot B} = c_F^{em}, \, V^{\sigma \cdot(E \times \nabla _r)} = c_s^{em} \, .
\end{equation}

With the help of the formalism developed in \cite{M1_Brambilla} we can describe the states in a quantum mechanical way using wave functions and compute the decay rate at NLO, i.e. at relative order $v^2$, from the Lagrangian (\ref{LagE1}). We obtain
\begin{equation}\label{E1_final}
 \Gamma _{n^3 P_J \rightarrow {n'}^3 S_1 \gamma} = \frac{4}{9} \, \alpha _{em} e_Q ^2 k_{\gamma}^3 I_3 ^2(n1 \rightarrow n'0) \left( 1 + R - \frac{k_{\gamma}^2}{60} \frac {I_5}{I_3} - \frac{k_{\gamma}}{6m} + \frac{k_{\gamma} (c_F^{em}-1)}{2m} \left[\frac{J(J+1)}{2} -2 \right] \right) \, ,
\end{equation}
where
\begin{equation}
  I_N \equiv \int _0 ^{\infty} dr  \, r^N R_{n'0} (r) R _{n1} (r) \, .
\end{equation}
$R$ contains all of the wave-function corrections due to the higher-order potentials mentioned in (\ref{V1})-(\ref{V3}), the relativistic correction of the kinetic energy, $-{\bf p}^4/4m^3$, and higher-order Fock state contributions due to intermediate color-octet states. In contrast to M1 transitions the latter ones do not vanish for E1 decays (see \cite{E1_Brambilla} for explicit expressions).

The expression (\ref{E1_final}) is also valid in the strongly coupled regime (without color-octet contributions in $R$), where $p \sim \Lambda_{QCD}$, since we made use of non-perturbative matching arguments and additional operators do not appear in this regime.

Compared to the results with the potential model calculation in \cite{Grotch} we find an equivalence between (\ref{E1_final}) and the corresponding formula there at the given order. However, our definite power counting allowed us to include all relativistic corrections systematically, in particular the color-octet contributions in the weak-coupling regime and the one coming from the potential $V_r ^{(1)}$. Both were missing in former approaches. Furthermore we can show that the anomalous magnetic moment $c_F^{em}-1 \sim \mathcal{O}(\alpha_s(m))$ is actually suppressed and does not lead to large non-perturbative contributions.

Without much effort one can extend the discussion to other processes like $n^1 P_1 \rightarrow {n'}^1 S_0 \gamma$ and $n^3 S_1 \rightarrow {n'}^3 P_J \gamma$ (see \cite{E1_Brambilla}), also for transitions between states with the same principal quantum number, where corrections $\sim k_\gamma$ are suppressed. 

\section{Phenomenological analysis}

Based on these results a phenomenological analysis for bottomonium and charmonium decays at relative order $v^2$ can be performed. For a complete analysis we need a parametrization of chromoelectric field correlators in the weak coupling regime. Since E1 transitions always involve excited states, we alse require the quarkonium potentials in the strong coupling regime. These have to be matched with the known short distance behaviour.

As a first approach we proceed as following: 
As a parametrization of the static potential at short distances we use the perturbative expression at 3 loop with leading ultrasoft resummation with the parameters given in \cite{alphas_det,Tormo} (see also references therein)\footnote{To obtain a convergent behaviour one has to perform a renormalon subtraction at a scale $\rho$ \cite{Pineda:renormalon}. In \cite{alphas_det} $\alpha_s(1/r)$ was expanded in terms of $\alpha_s(\rho)$ to get a more stable behaviour. For short distances, $r<0.14r_0$, large logarithms yield an unreasonable shape of the potential, therefore we expand $\alpha_s(\rho)$ in terms of $\alpha_s(1/r)$ in this regime.} derived from a matching of the static energy to unquenched lattice data \cite{lattice_unquenched}. For large distances we use the string potential $V_{\rm{string}}=-\pi/12r+\sigma r+C$ \cite{Luscher} matched to lattice simulations \cite{lattice_unquenched} at $r=1.5 r_0$ (where $r_0$ is the Sommer scale). These two potentials merge together smoothly at $r \sim 0.8 r_0$. To 
obtain the leading order wave functions we solve the Schr\"odinger equation with the static potential numerically using the \emph{Mathematica} program \emph{ schrodinger.m} \cite{Schrodinger_eq}. We fix the charm and bottom mass in our scheme a posteriori by matching the obtained masses for the $J/\psi$ and $\Upsilon(1S)$ with the physical ones.

Concerning relativistic corrections the main contribution arises from wave function corrections due to subleading potentials. For short distances, here for $r<(2 \, \rm{GeV})^{-1}$, we apply perturbative results at LL, wheras for long distances, $r>(2 \, \rm{GeV})^{-1}$, we use parametrizations from quenched lattice results \cite{Koma_m2_SD,Koma_m2_SI,Koma_new} as a non-perturbative input.\footnote{As far as available we use fits with parametrizations based on calculations from the string model \cite{PerezNadal}. $V_r^{(2)}$ (so far undetermined) and $V_{S^2}^{(2)}$ (small) are set to 0 in the non-perturbative regime.} Future approaches should aim for smooth transitions between these two regimes. We neglect color octet effects, which cannot be determined from current lattice simulations.

The results of this computation are given in table \ref{table_bottomonium} for bottomonium and in table \ref{table_charmonium} for charmonium decays. We do not include decays with $k_\gamma \gtrsim \langle p \rangle$, where our power counting is assumed to break down. We see that the relativistic corrections lower the decay rates considerably, by 10-30\% for bottomonium and by 20-60\% for charmonium, which is especially striking for the decays $h_c (1P) \rightarrow \eta_c(1S) \gamma$ and $\psi(2S) \rightarrow \chi_{c0}(1P) \gamma$. The wave function correction due to the potential $V_r^{(1)}$ yields particularly large contributions. 
The expansion works much better for bottomonium, since the average relative velocity is smaller ($v_b^2 \sim 0.1$, $v_c^2 \sim 0.3$). We estimate the uncertainty of our NLO result to be of order 10\% for bottomonium and of order 30\% for charmonium. This is on the one hand the generic size of a correction at $\mathcal{O}(v^2)$, which can arise from color-octet effects or a systematic error in the treatment of the subleading potentials. On the other hand this is also a conservative measure for the total higher order effects, which are supposed to be suppressed by $v$ compared to the total NLO corrections. 

Comparing our values with the potential model calculations in \cite{Grotch} (for scalar and vector confining potential) we tend to get slightly larger values for bottomonium and slightly smaller values for charmonium. Within our uncertainties we stay consistent with observations, provided both the branching fraction and the total decay rate have been measured.

We emphasize that except for the masses of the $J/\psi$ and $\Upsilon(1S)$ and the photon energies no experimental input has been used. Additional input and a thorough investigation of the fine splitting effects in the spectrum would improve the predictive power of our results. This analysis should be repeated, when unquenched results for the required chromoelectric field correlators and especially for the subleading potentials become available, with emphasis on a proper connection to perturbative results, maybe at higher order (NLL), and a more elaborate uncertainty estimate.

\begin{table}
\centering
   \begin{tabular}{c|c|c|c|c}
	process & $\Gamma_{\rm{pNRQCD}}^{\rm{LO}}$/keV & $\Gamma_{\rm{pNRQCD}}^{\rm{NLO}}$/keV & $\Gamma_{\rm{mod}}^{\cite{Grotch}}$/keV & $\Gamma_{\rm{exp}}^{\textrm{PDG}}$/keV \\
	\hline
	$\chi_{b0}(1P) \rightarrow \Upsilon(1S) \gamma$ & 31.8 & 29.7 $\pm$ 3.1 & 25.7-27.0 & - \\ 
	$\chi_{b1}(1P) \rightarrow \Upsilon(1S) \gamma$ & 40.3 & 35.8 $\pm$ 4.0 & 29.8-31.2 & - \\ 
	$\chi_{b2}(1P) \rightarrow \Upsilon(1S) \gamma$ & 45.9 & 40.6 $\pm$ 4.6 & 33.0-34.2 & - \\ 
	$h_b(1P) \rightarrow \eta_b(1S) \gamma$ & 60.8 & 44.3 $\pm$ 6.1 & - & - \\
	$\Upsilon(2S) \rightarrow \chi_{b0}(1P) \gamma$ & 1.52 & 1.13 $\pm$ 0.15 & 0.72-0.73 & 1.22 $\pm$ 0.16 \\
	$\Upsilon(2S) \rightarrow \chi_{b1}(1P) \gamma$ & 2.26 & 1.94 $\pm$ 0.23 & 1.62-1.65 & 2.21 $\pm$ 0.22 \\
	$\Upsilon(2S) \rightarrow \chi_{b2}(1P) \gamma$ & 2.34 & 2.19 $\pm$ 0.23 & 1.84-1.93 & 2.29 $\pm$ 0.22 \\
	$\chi_{b0}(2P) \rightarrow \Upsilon(2S) \gamma$ & 12.6 & 13.0 $\pm$ 1.3 & 10.6-11.4 & - \\
	$\chi_{b1}(2P) \rightarrow \Upsilon(2S) \gamma$ & 17.1 & 16.3 $\pm$ 1.7 & 11.9-12.5 & - \\
	$\chi_{b2}(2P) \rightarrow \Upsilon(2S) \gamma$ & 20.4 & 18.1 $\pm$ 2.0 & 12.9-13.1 & - \\
	$\Upsilon(3S) \rightarrow \chi_{b0}(2P) \gamma$ & 1.44 & 1.05 $\pm$ 0.14 & 1.07-1.09 & 1.20 $\pm$ 0.16 \\
	$\Upsilon(3S) \rightarrow \chi_{b1}(2P) \gamma$ & 2.38 & 2.05 $\pm$ 0.24 & 2.15-2.24 & 2.56 $\pm$ 0.34 \\
	$\Upsilon(3S) \rightarrow \chi_{b2}(2P) \gamma$ & 2.53 & 2.35 $\pm$ 0.25 & 2.29-2.44 & 2.66 $\pm$ 0.41 
  \end{tabular} 
  \caption{E1 decay rates for bottomonium. Our pNRQCD results compared to a potential model calculation \cite{Grotch} and and the current PDG values \cite{PDG2012}. LO denotes the result obtained without relativistic corrections, NLO indicates the result up to $\mathcal{O}(v^2)$ neglecting color-octet effects in the weak-coupling regime and non-perturbative contributions to $V_r^{(2)}$. The error estimates give the generic size of one $\mathcal{O}(v^2)$ correction as well as an estimate for the sum of all corrections at $\mathcal{O}(v^3)$. For $h_b(1P) \rightarrow \eta_b(1S) \gamma$ we have taken $m_{\eta_b(1S)}=9402$ GeV \cite{eta_b} to determine the photon energy.}
  \label{table_bottomonium}
\end{table}

\begin{table}
\centering
 \begin{tabular}{c|c|c|c|c}
	process & $\Gamma_{\rm{pNRQCD}}^{\textrm{LO}}$/keV & $\Gamma_{\rm{pNRQCD}}^{\textrm{NLO}}$/keV & $\Gamma_{\rm{mod}}^{\cite{Grotch}}$/keV & $\Gamma_{\rm{exp}}^{\textrm{PDG}}$/keV \\
	\hline
	$\chi_{c0}(1P) \rightarrow J/\psi \gamma$ & 199 & 158 $\pm$ 60 & 162-183  & 122 $\pm$ 11 \\
	$\chi_{c1}(1P) \rightarrow J/\psi \gamma$ & 421 & 302 $\pm$ 126 & 340-363  & 296 $\pm$ 22  \\
	$\chi_{c2}(1P) \rightarrow J/\psi \gamma$ & 568 & 415 $\pm$ 170 & 413-464 & 386 $\pm$ 27 \\
        $h_c (1P) \rightarrow \eta_c(1S) \gamma$ & 909 & 447 $\pm$ 272 & - & $<$600  \\
        $\psi(2S) \rightarrow \chi_{c0}(1P) \gamma$ & 53.6 & 21.4 $\pm$ 16.1 & 26.0-40.3  & 29.4 $\pm$ 1.3 \\
	$\psi(2S) \rightarrow \chi_{c1}(1P) \gamma$ & 45.2 & 30.7 $\pm$ 13.6 & 28.3-37.3  & 28.0 $\pm$ 1.5  \\
	$\psi(2S) \rightarrow \chi_{c2}(1P) \gamma$ & 31.6 & 25.6 $\pm$ 9.5 & 17.5-22.7 & 26.5 $\pm$ 1.3 \\
        $\eta_c(2S) \rightarrow h_c(1P) \gamma$ & 38.1 & 31.0 $\pm$ 11.4 & - & -
  \end{tabular} 
  \caption{E1 decay rates for charmonium.  Our pNRQCD results at LO, NLO (including error estimate) compared to a potential model calculation \cite{Grotch} and the current PDG values \cite{PDG2012}.}
  \label{table_charmonium}
\end{table}

\subsection*{Acknowledgements}
  I would like to thank Nora Brambilla and Antonio Vairo for the collaboration on this work.

\end{document}